\documentstyle[epsfig,epsf]{article}
\begin{document}

\begin{center}
{\Large {\bf BULK ETCH RATE MEASUREMENTS AND CALIBRATIONS OF PLASTIC NUCLEAR TRACK DETECTORS}}
\end{center}

\vskip .7 cm

\begin{center}
S. Balestra$^1$,  M. Cozzi$^1$,  G. Giacomelli$^1$,  R. Giacomelli$^1$,  
M. Giorgini$^1$,  
A. Kumar$^{1,2}$,     
G. Mandrioli$^1$,   S. Manzoor$^{1,3}$,  A. R. Margiotta$^1$,  
E. Medinaceli$^1$,  
L. Patrizii$^1$,  V. Popa$^{1,4}$,  
I.E. Qureshi$^3$, M.A. Rana$^3$, G. Sirri$^1$, M. Spurio$^1$, 
V. Togo$^1$ and C. Valieri$^1$
 \par~\par

{\it  1. Phys. Dept. of the University of Bologna and INFN, Sezione di 
Bologna, Viale C. Berti Pichat 6/2, I-40127 Bologna, Italy \\ 
2. Dept. Of Physics, Sant Longowal Institute of Eng. and Tech., Longowal 
148 106 India \\
3. PRD, PINSTECH, P.O. Nilore, Islamabad, Pakistan \\
4. Institute of Space Sciences, Bucharest R-077125, Romania} 

\par~\par

\vskip .7 cm
{\large \bf Abstract}\par
\end{center}

{\normalsize New calibrations  of CR39 and Makrofol nuclear track detectors 
have been obtained using 158 A GeV $Pb^{82+}$ and $In^{49+}$ ions; a new 
method for the bulk etch rate determination, using both cone height and base 
diameter measurements was developed. The CR39 charge resolution based on the 
etch-pit base area measurement is adequate to identify nuclear fragments in 
the interval $7 \leq Z/\beta \leq 49$.  For CR39 the detection threshold is at 
REL $\sim 50$ MeV cm$^2$ g$^{-1}$, corresponding to a nuclear fragment with 
Z/$\beta \sim$ 7.  
Base cone area distributions for Makrofol foils exposed to $Pb^{82+}$ ions 
have 
shown for the first time all peaks due to nuclear fragments with $Z > 50$; 
the distribution of the etched cone heights shows well separated individual 
peaks for Z/$\beta$ = 78 $\div$ 83 (charge pickup). The Makrofol detection 
threshold is 
at REL $\sim 2700$ MeV cm$^2$ g$^{-1}$, corresponding to a nuclear fragment 
with Z/$\beta \sim$ 50.\\
 

}

\vspace{5mm}

\large
\section{Introduction}\label{sec:intro}Nuclear Track Detectors (NTDs) are
 employed in several scientific and technological applications [1,2].
 The most sensitive NTD is the isotropic poly-allyl-diglycol carbonate polymer,
 commercially known as CR39$^{\small{\textregistered}}$; Makrofol$^{\small{\textregistered}}$/Lexan$^{\small{\textregistered}}$ polycarbonates are also 
largely employed. More than 4000 $m^2$ of CR39 detectors were used in the  MACRO 
and SLIM experiments devoted to the search for new massive particles in the 
cosmic radiation (magnetic monopoles, nuclearites, q-balls) [3,6]. 
Several experiments are going on in different fields which require an accurate detector calibration [7-8].\par
The damaged trail (called ``latent track'') produced by an ionizing particle 
can be made visible with an optical microscope through a chemical etching 
process in aqueous solution of  either NaOH or  KOH at a proper concentration 
and temperature. The latent track develops into a conical-shaped etch-pit, 
when the etching velocity along the particle trajectory ($v_T$) is larger than 
the one for the bulk etching of the material ( $v_B$), Fig. 1 [1]. 
The addition of ethyl 
alcohol in the etchant speeds up the etching process, improves the post-etched 
surface quality of CR39 and Makrofol, but raises their detection thresholds.\par
	For particles with constant energy loss, the  etch-rate ratio 
$p = v_T/v_B$, may be determined by measuring either the etch-pit surface area 
or the 
etch-pit height and  the bulk etch rate,  $v_B$. Two methods have been used 
to 
determine  $v_B$: the most common one is based on the measurement of the 
detector 
thickness removed during etching. The method is affected by a systematic 
error of a few $\%$ [9].\par
	The  main aim of this paper is to analyze a new method of 
measurement of the bulk-etch rate in CR39 and in Makrofol, and to obtain new 
calibrations for these detectors.

\section{Experimental}A stack composed of Makrofol and CR39 foils of size 
11.5 x 11.5 cm$^2$ with a 1 cm 
thick lead target was 
exposed to 158 A GeV $Pb^{82+}$ ions in 1996 (Pb96); a second stack with a 
1 cm thick aluminium 
target was exposed to 158 A GeV $In^{49+}$ ions in 2003 (In03); 
both exposures were performed at the CERN-SPS, at normal incidence and a total 
ion density of $\sim 2000 / cm^2$. The CR39 polymer sheets used in the present 
experiment were manufactured by Intercast Europe Co., Parma, Italy using a 
specially designed line of production [10]. The Makrofol detectors were 
manufactured by Bayer A.G., Germany. The Makrofol thickness is 500 $\mu m$, 
the 
CR39 thickness is either 700 $\mu m$ or 1400 $\mu m$; all  detector sheets 
were covered 
by a 30 $\mu m$ plastic film to protect them from exposure to ambient radon; the protective layers were removed before etching. The 
detector foils downstream of the target recorded the beam ions as well as 
their nuclear  fragments. \par 
After exposures, two CR39 foils (In03 stack) and two 
Makrofol foils, (Pb96 stack), located after the target were etched in 6 N 
NaOH + 1$\%$ ethyl alcohol at 70 $^\circ C$ for 40 h and 6 N KOH + 20 $\%$ 
ethyl 
alcohol at 50 $^\circ C$ for 8 h, respectively. The etching was performed in a 
stainless steel tank equipped with internal thermo-resistances and a motorized 
stirring head. The temperature was stable to within $\pm 0.1$ $^\circ C$. 
In order to keep 
homogeneous the solution and to avoid that etched products deposit on the 
detector surfaces, a continuous stirring was applied during etching.\par
	For CR39 detectors, etch-pit base diameters and heights of In ions and 
their fragments were  measured with a Leica optical microscope. In Makrofol, 
Pb ions and their high Z fragments made through-holes in the detector sheets; 
thus the cone length $L_e$ was measured only for high Z fragment tracks that 
have 
sharp etch-cone tips (no holes). Nuclear fragments with charges 
$78 \leq Z \leq 82$ 
were identified by etching another Makrofol sheet from the same stack in the 
same conditions for only 5 hours.

\subsection{``Standard" measurement of $v_B$}
	As already recalled, the standard determination of $v_B$ 
is based on the measurement of the thickness of the detector at different 
etching times. The thickness is measured with an electronic micrometer of 1 
$\mu m$ accuracy in 25 positions on the detector foil. The average bulk-etch 
velocity is $v_B = \Delta x / 2 \Delta t$, where $\Delta x$ is the mean 
thickness difference after a $\Delta t$ etching time. 
For CR39, at etching times shorter than 10 hours the thickness is affected by 
detector swelling [11-13]. The bulk etching rate must be determined by a 
linear fit of $\Delta x$ vs $\Delta t$ for etching times longer than 10 hours. 
For Makrofol no 
significant swelling effect was observed.

\subsection{The bulk etch rate from the cone height and base diameter measurements}
	For relativistic charged particles the track etch rate $v_T$ can be 
considered constant. For normally incident particles, the measurable 
quantities are the  cone base diameter D, and the height $L_e$, see Fig. 1. 
$L_e$ is obtained by multiplying the measured cone height by the refractive index n of the etched detector material; n is obtained from the ratio of the actual thickness (measured with an electronic micrometer with a precision of 1 
$\mu m$) to the apparent thickness measured with an optical microscope 
(precision of 1 $\mu m$) [$n_{CR39} = 1.55 \pm 0.01$; 
$n_{Makrofol} = 1.69 \pm 0.01$].\par
The following relations  hold: 

\begin{equation}
L_e = (v_T-v_B)t 
\end{equation}

\begin{equation}
D= 2 v_B t \sqrt { \frac{(v_T-v_B)}{(v_T+v_B)} }    
\end{equation}	

From the above relations, the following quadratic equation in  $v_B$ is 
obtained

\begin{equation}
\left(\frac{L_e}{t}\right) v_B^2 - \left(\frac{D^2}{2t^2}\right) 
v_B-\left(\frac{D^2 L_e}{4t^3}\right) =0
\end{equation}

The real solution for $v_B$ is

\begin{equation}
v_B= \frac{D^2}{4t L_e} \left[ 1+ \sqrt{ 1+ \frac{4L_e^2}{D^2}} \right]
\end{equation}

From equation 1 the track etch rate $v_T$ can be written as

\begin{equation}
v_T=v_B + \frac{L_e}{t}
\end{equation}

and from equations 1 and 2, the reduced etch rate follows

\begin{equation}
p= \left(\frac{v_T}{v_B} \right) = 1 + \frac{L_e}{v_B t} = 
\frac{1+(D/2v_B t)^2}{1-(D/2v_B t)^2}
\end{equation}

	We may thus determine the bulk etch rate $v_B$ and the reduced 
etch-rate p by measuring the track parameters Le (measured with a precision 
of $\sim$ 1 $\mu m$) and D (precision of 0.5 $\mu m$).\par
Relations (4-6) were tested with relativistic Pb and In ions and their nuclear 
fragments. We selected only tracks for which precise measurements of the cone 
height and diameter could be performed (for example we cannot measure precisely
 the 
track cone heights for low Z fragments). Then, using equation (4) we computed 
the 
bulk-etch rate for CR39 and Makrofol. Batches of measurements were made by 
different operators, and the average $v_B$'s and their statistical standard 
deviations were computed, see Table 1. By this method we obtain $v_B$  values 
with accuracies of $\pm$ 0.01 - 0.05 $\mu m$/h. The $v_B$ values obtained for 
the same 
foils using detector thickness measurements are also given.\par
Notice that we can effectively make measurements with the available measuring 
instruments using the new method only for intermediate z-values because at low 
and high z we do not have enough precision for measuring $L_e$ and D, 
respectively.

\section{Calibrations}
In ref. [14] it was shown that the reduced etch rate p of CR39 is a unique 
function of the particle Restricted Energy Loss (REL). In the following 
sections new calibration data are reported for CR39 and Makrofol based on 
the new determination of the bulk etch rate.

\subsection{Calibration of the CR39 detector}
Etch-pit base areas were measured with the Elbek automatic image analyzer 
system [15]. Fig. 2 shows the etch-pit base area distribution for Indium ions 
and their fragments in CR39; averages were computed from measurements made on 
the "front sides" of two detector sheets. The peaks are well separated from 
$Z/ \beta \sim$ 7 to 45; the charge resolution for the average of two 
measurements is $\sigma_Z \sim 0.13e$ at 
$Z/ \beta \sim 15$. The charge 
resolution close to the Indium peak (Z $\sim$ 49) can be improved by measuring the 
heights of the etch pit cones [16]. The heights of 1000 etch-cones with 
diameter larger than 48 $\mu m$ (corresponding to nuclear fragments with 
$Z > 45$) 
were measured with an accuracy of ñ $\pm 1$ $\mu m$ with a Leica microscope coupled to a 
CCD camera and a video monitor. The corresponding distribution is shown in the 
inset in Fig. 2; each of the 4 peaks is well separated from the others, 
and a charge 
can  be assigned to each one.\par\par 
The charge resolution for single measurements of different nuclear 
fragments is given in Table 2; it is 
$\sigma_Z \sim 0.22e$ at $Z/\beta \sim 48$. The charge 
resolutions may be computed using the relations [17] 

\begin{equation}
\sigma_Z = \frac{\sigma_A}{\delta A / \delta Z} ~~ 
, ~~ \sigma_Z = \frac{\sigma_{L_e}}{\delta L_e / \delta Z}
\end{equation}
							        
where $\sigma_A$ and $\sigma_{L_e}$ are the standard deviations; A is 
the mean base area and $L_e$ 
the mean height of the etched cones.

\subsection{Calibration of the Makrofol detector}
Fig. 3 shows the base area distribution for the average of 2 measurements of 
Pb ions and their fragments in 
Makrofol; averages were computed from measurements on the front sides of two 
detector foils. The peaks are well separated from $Z/ \beta = 51$ to 
$\sim 77$ (the charge 
resolution is $\sigma_Z \sim 0.18e$ at $Z/ \beta \sim 55$. The charge 
resolution close to the Pb 
peak (Z = 82) was improved by measuring the heights of the etch pit cones. The 
heights of 4000 etch cones with base diameters larger than 47 $\mu m$ were 
measured; the corresponding distribution is shown in the inset in Fig. 3; 
each peak is well separated from the others, and a charge was assigned to 
every peak. The charge resolution for single measurements of different nuclear fragments is  $\sigma_Z \sim 0.18e$ at $Z/ \beta \sim 81$, see Table 2). 
Notice the presence of the 
Z = 83 peak from a charge pick-up reaction.\par
	For each detected nuclear fragment from Z = 7 to 48 and Indium ions 
(Z = 49) we computed the REL and the reduced etch rate p = $v_T / v_B$ 
using eq. (6).  p versus REL for CR39 is plotted in Fig. 4; the CR39 detection 
threshold is 
at REL $\sim 50$ MeV cm$^2$ g$^{-1}$ (corresponding to a relativistic nuclear 
fragment with $Z \sim$ 7.\par 
The same procedure was applied to Makrofol (Fig. 5). The Makrofol detection threshold is at REL $\sim 2700$ MeV cm$^2$ g$^{-1}$, corresponding to a nuclear 
fragment with $Z/ \beta \sim 50$.\par 
As evident in Figs. 4 and 5 the reduced etch-rate p is a non-linear function of REL. For example for CR39 up to REL $\sim 500$ MeV cm$^2$ g$^{-1}$, p changes 
slowly with REL, while a rapid increase is observed at larger values.\par 
The changes $\delta D$ and $\delta L_e$ with respect to $\delta p$  can be 
obtained by differentiating eqs. 1 and 2 with respect to p [18]:

\begin{equation}
\frac {\delta D}{\delta p} = 2 v_B t \sqrt {\frac{p+1}{p-1}} \frac{1}{(p+1)^2}
\end{equation}

\begin{equation}
\frac{\delta L_e}{\delta p} = v_B t 
\end{equation}

At large p values, $\delta D / \delta p$ is smaller with respect to 
$\delta L_e / \delta p$; for high REL it is difficult to obtain a charge 
resolution $<$ 1e. From Figs. 2 and 3 it is seen that for high $Z/ \beta$, the 
base area distribution does not give well separated peaks, while by cone 
height measurements the peaks are well separated (see the insets).

\section{Discussion and Conclusions}
The "new  method" for measuring the bulk etch rate for intermediate or high z values yields slightly smaller 
uncertainties than the "standard method" (change in thickness). This comes 
from the use at the same time of both cone heights and base diameter 
measurements of tracks.\par 
The values obtained here by the two methods are in reasonable agreement; the 
differences may arise from small systematic uncertainties affecting 
the thickness measurement of a sheet and the cone height and diameter of 
etch-pits.\par
Calibration data were obtained with $In^{49+}$ ions of 158 A GeV for CR39 and 
with $Pb^{82+}$ ions of 158 A GeV for the Makrofol detectors. 
Well separated peaks for 
the primary ions and for their fragments are observed in Figs. 2 and 3. At 
low Z/$\beta$ the measurements of the base area cones are adequate, while at 
high Z/$\beta$ the measurements of the cone heights are more useful. All the 
peaks are well separated in CR39 (In03) for Z/$\beta \sim$ 7 - 49 and in Makrofol 
(Pb96) for Z/$\beta \sim$ 52 - 83; a charge value may be assigned to each peak 
(for these exposures $\beta \sim 1$).\par 
The reduced etch rate p (computed with the new method) plotted versus 
REL covers a large Z/$\beta$ range for both detectors, Fig. 4 and 5.

\section{Acknowledgements}
We thank the CERN SPS staff for the Pb and In beam exposures. 
We acknowledge many colleagues for their cooperation and technical advice. 
We gratefully acknowledge the contribution of our technical staff, 
in particular  E. Bottazzi, L. Degli Esposti and G. Grandi. 
We thank INFN and ICTP for providing fellowships and grants to non-Italian 
citizens.

\begin{table}[!ht]
\begin{center}
\caption{Bulk etch rates $v_B$ for CR39 and Makrofol NTDs obtained with the 
new and the standard methods using 25 measurements for each final data point. 
The errors are statistical standard deviations of the mean. The different 
values of $v_B$ for Pb96 in rows 3 and 4 are due to the different etch temperatures. }
\vspace{1 cm}

\begin{tabular}{|c|c|c|c|c|}
\hline
Detector & Z Range & Etching Conditions & $v_B$      & $v_B$ \\
(beam)   &         &                    & New Method & Standard Method \\
\hline
CR39  & 44 - 49 & 6N NaOH  & 1.25 $\pm ~0.01 \mu$m/h & 1.15 $\pm ~0.03 \mu$m/h \\
(In03)&        & +1 $\%$ alcohol, 70$^\circ$C, 40 h &  & \\
\hline
CR39  & 75 - 80 & 6N NaOH 70$^\circ$C, 30 h  & 1.10 $\pm ~0.02 \mu$m/h & 1.15 
$\pm ~0.03 \mu$m/h \\
(Pb96)&        & &  & \\
\hline
CR39  & 78 - 82 & 6N NaOH 45$^\circ$C, 268 h  & 0.16 $\pm ~0.01 \mu$m/h & 0.17 
$\pm ~0.03 \mu$m/h \\
(Pb96)&        & &  & \\
\hline
Makrofol  & 75 - 78 & 6N KOH  & 3.44 $\pm ~0.05 \mu$m/h & 3.52 $\pm ~0.13 \mu$m/h \\
(Pb96)&        & +20 $\%$ alcohol, 50$^\circ$C, 8 h &  & \\
\hline
\end{tabular}
\end{center}
\end{table}

\begin{table}[!ht]
\begin{center}
\caption{Assigned charges and estimated charge resolutions (for the average of 2 base cone area measurements or a cone height measurement, see text) 
for $In^{49+}$ and 
$Pb^{82+}$ ions and their fragments in CR39 and Makrofol detectors. 
In CR39 the 
uncertainty in the charge resolution is  about $\pm$ 0.02 (by the 
measurements of the etch-pit base areas, rows 3 to 8) and $\pm$ 0.04 (by cone 
height measurements, last row). In Makrofol it is  $\pm$ 0.03 }
\vspace{1 cm}
\begin{tabular}{|c|c|c|c|}
\hline
\multicolumn{2}{|c|}{\bf CR39 Detector} &
\multicolumn{2}{|c|}{\bf Makrofol Detector} \\ 
\hline
Charge & Charge Resolution & Charge & Charge Resolution \\
\hline
Z=8 to 11 & 0.12e & Z=51 to 58 & 0.18e \\
\hline
Z=12 to 21 & 0.13e & Z=51 to 66 & 0.19e \\ 
\hline
Z=22 to 31 & 0.16e & Z=59 to 66 & 0.21e \\
\hline
Z=32 to 41 & 0.20e & Z=59 to 69 & 0.22e \\
\hline
Z=32 to 45 & 0.22e & Z=70 to 74 & 0.31e \\
\hline
Z=42 to 49 & 0.28e & Z=75 to 77 & 0.37e \\
\hline
\hline
Z=46 to 49 & 0.22e & Z=79 to 82 & 0.18e \\
by cone height & & by cone height & \\
\hline

\end{tabular}\label{tab:hi-res}
\end{center}
\end{table}

\begin{figure}[!ht]
\begin{center}
\mbox{\epsfig{figure=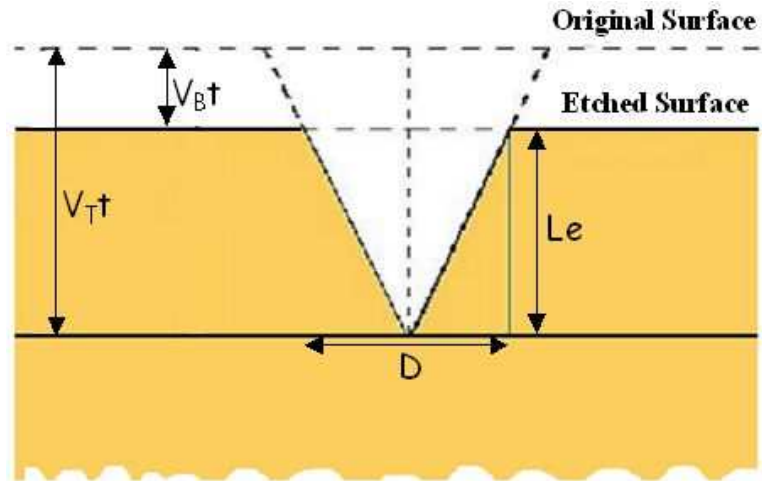,height=6.5cm}}
\caption{Sketch of an "etched track" for a normally incident ion in a nuclear 
track detector. }
\label{fig:1}
\end{center}
\end{figure}

\begin{figure}
\begin{center}
\mbox{\epsfig{figure=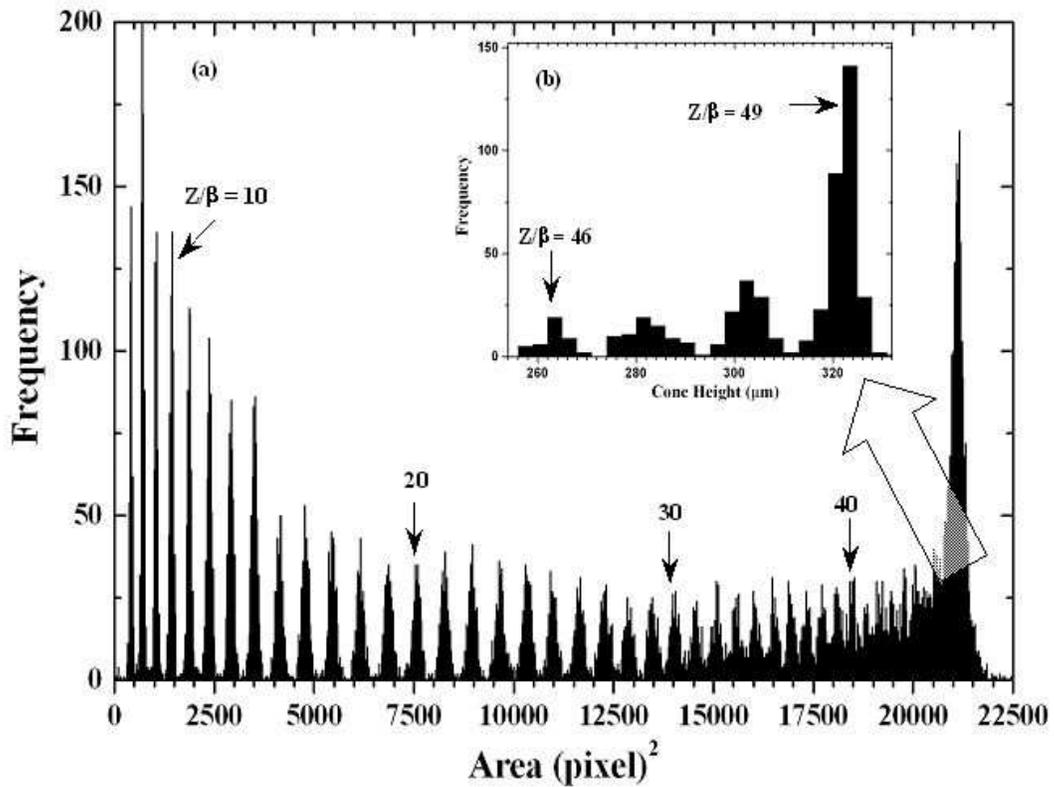,height=10.5cm}}
\caption{(a) Base area distribution of etched cones in CR39 from  158 A GeV 
$In^{49+}$ ions and their fragments (averages of 2 front face measurements); (b) cone height distribution for $46 \leq Z/ \beta \leq 49$. Etching conditions: 
6 N NaOH + 1 $\%$ ethyl alcohol, $70^{\circ}$ C, 40 h.}
\label{fig:2}
\end{center}
\end{figure}

\begin{figure}
\begin{center}
\mbox{\epsfig{figure=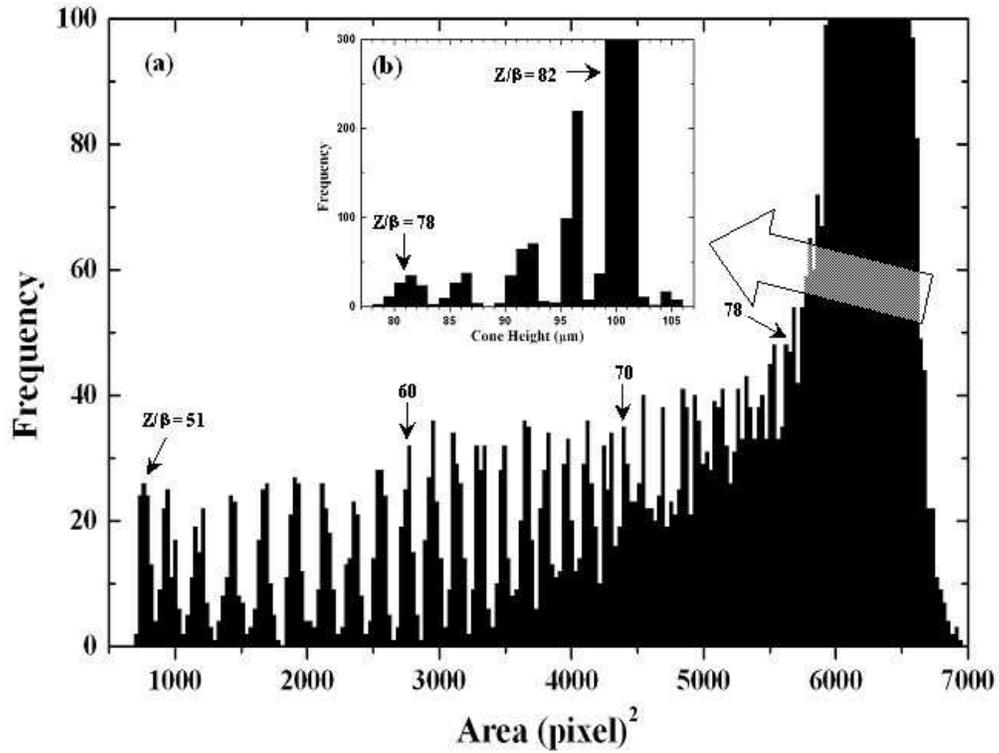,height=10cm}}
\caption{(a) Base area distribution of  etched cones in Makrofol from 158 A 
GeV $Pb^{82+}$ ions and their fragments (averages of 2 front face measurements); (b) cone height distribution for $78 \leq Z/ \beta \leq 83$ (pick up). 
Etching conditions: 6 N KOH + 20 $\%$ ethyl alcohol, $50^{\circ}$ C, 8 h. }
\label{fig:3}
\end{center}
\end{figure}

\begin{figure}
\begin{center}
\mbox{\epsfig{figure=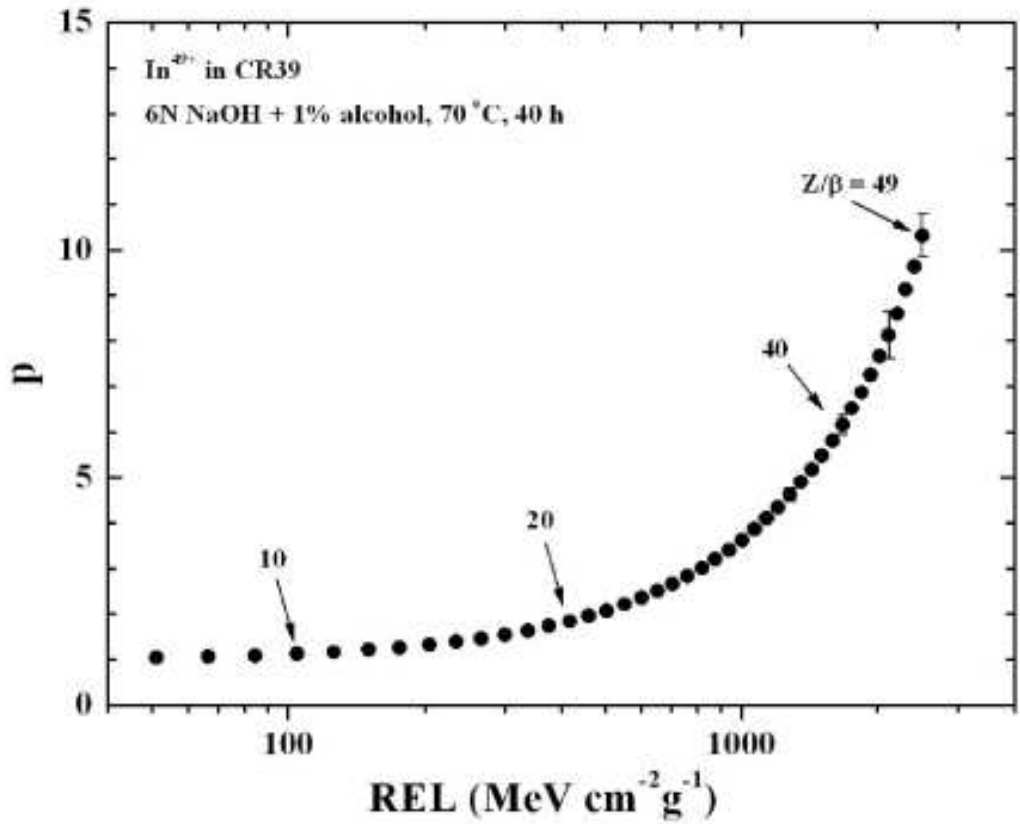,height=13cm}}
\caption{p vs. REL for CR39  exposed to relativistic Indium ions using $v_B$ 
evaluated with the new method. Typical statistical standard deviations are 
shown at Z/$\beta$ = 40, 45, 49; for Z/$\beta \leq 37$ the errors are inside the 
black points. }
\label{fig:4}
\end{center}
\end{figure}

\begin{figure}
\begin{center}
\mbox{\epsfig{figure=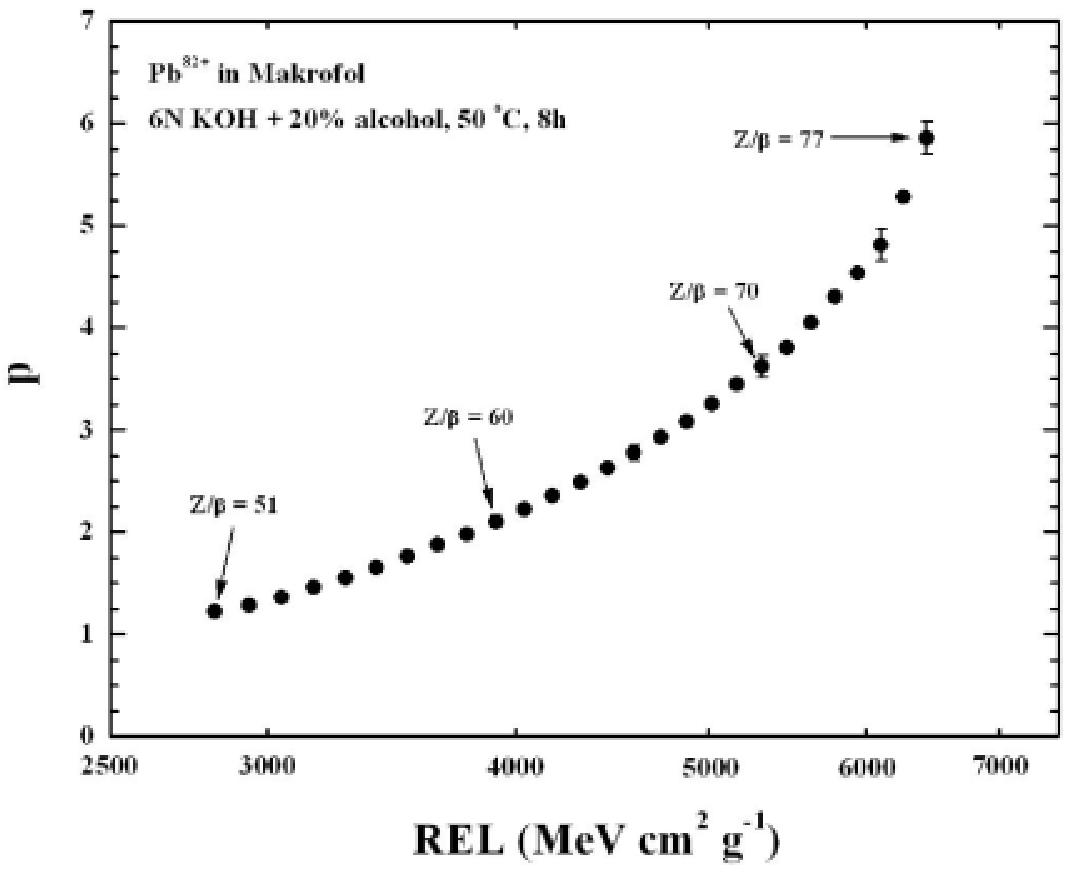,height=13cm}}
\caption{p vs. REL for the Makrofol detector exposed to relativistic Pb ions 
using $v_B$ evaluated with the new method. Typical statistical standard 
deviations are shown at Z/$\beta$ = 70, 75, 77; for Z/$\beta \leq 67$ the 
errors are inside the black points. }
\label{fig:5}
\end{center}
\end{figure}


\begin{thebibliography}{20}

\bibitem{1} R.L Fleischer, P.B. Price, R.M. Walker, Nuclear Tracks in Solids, 
University of California Press, California, 1975.

\bibitem{2} S.A. Durrani, R.K. Bull, Solid State Nuclear Track Detection, 
Pergamon Press, Oxford, 1987.

\bibitem{3} M. Ambrosio et al., Eur. Phys. J. C 25 (2002) 511; hep-ex/0207020.

\bibitem{4} S. Cecchini et al., Radiat. Meas. 40 (2005) 405; hep-ex/0508043; 
hep-ex/ 0503003; hep-ex/0506075.

\bibitem{5} J. Derkaoui, G. Giacomelli, T. Lari, G. Mandrioli, M. Ouchrif, 
L. Patrizii, V. Popa, Astropart. Phys. 10 (1999) 339.

\bibitem{6} S. Cecchini, T. Chiarusi, G. Giacomelli, A. Kumar, L. Patrizii, 
Proc. of the 16th PAC-European Rocket $\&$ Balloon Programmes $\&$ Related Research, 
(2003), SP-530, astro-ph/0510717.

\bibitem{7} Y. Uchihori et al. (ICCHIBAN Collaboration), J. Radiat. Res. 
(Tokyo) 43 (2002) Suppl:S81-5.

\bibitem{8} S. Kodaira, N. Hasebe, T. Doke, A. Kitagawa, H. Kitamura, S. Sato, 
Y. Uchihori, N. Yasuda, K. Ogura, H. Tawara, Japanese J. of Appl. Phys. 43 
(2004) 6358.

\bibitem{9} S. Cecchini, G. Giacomelli, M. Giorgini, L. Patrizii, P. Serra, 
Radiat. Meas. 34 (2001) 55.

\bibitem{10} L. Patrizii et al., Nucl. Tracks Radiat. Meas. 19 (1991) 641.

\bibitem{11} S. Ahlen et al., Nucl. Instrum. Meth. A 324 (1993) 337.

\bibitem{12} A. Kumar, R. Prasad, Nucl. Instrum. Meth. B 119 (1996) 515.

\bibitem{13} F. Malik, E.U. Khan, I.E. Qureshi, S.N. Husaini, M. Sajid, 
S. Karim, K. Jamil, Radiat. Meas. 35 (2002) 301.

\bibitem{14} S. Cecchini et al., Il Nuovo Cimento, 109 A (1996) 1119.

\bibitem{15} A. Noll, G. Rusch, H. Rocher, J. Dreute, W. Heinrich, 
Nucl. Tracks Radiat. Meas. 15 (1988) 265.

\bibitem{16} G. Giacomelli, M. Giorgini, G. Mandrioli, S. Manzoor, 
L. Patrizii, V. Popa, P. Serra, V. Togo, E.C. Vilela, Nucl. Instrum. 
Meth. A 411 (1998) 41.

\bibitem{17} S. Cecchini, H. Dekhissi, G. Giacomelli, E. Katsavounidis, 
A.R. Margiotta, L. Patrizii, F. Predieri, P. Serra, M. Spurio, Astropart. 
Phys. 1 (1993) 369.

\bibitem{18} I.E. Qureshi, M.I. Shahzad, M.T. Javed, S. Manzoor, G. Sher, 
F. Aleem, H.A. Khan, Radiat. Meas. 40 (2005) 437.

\end{thebibliography}
\end{document}